\documentclass[aps,prx,twocolumn,showpacs,floatfix,nobibnotes,nofootinbib,superscriptaddress,amsmath,amssymb,longbibliography]{revtex4-1}
\usepackage[english]{babel}
\usepackage[colorlinks,urlcolor={blue}]{hyperref}
\usepackage{eurosym}
\usepackage[usenames]{color}
\usepackage{graphicx}
\usepackage{amsmath}
\usepackage{amsfonts}
\usepackage{amssymb}
\usepackage{mathrsfs}
\usepackage{bm}
\usepackage{verbatim}
\usepackage{ulem}

\usepackage[textsize=small]{todonotes}

\newcommand{\beq}{\begin{equation}}
\newcommand{\eeq}{\end{equation}}
\newcommand{\bea}{\begin{eqnarray}}
\newcommand{\eea}{\end{eqnarray}}

\begin{document}

\title{Nuclear Physics under the low-energy, high intensity frontier}

\author{C.-J.~Yang}
\affiliation{Extreme Light Infrastructure - Nuclear Physics, "Horia Hulubei" National Institute for R\&D in Physics and Nuclear Engineering, 30 Reactorului Street, 077125 Magurele, Romania}
\email{chieh.jen@eli-np.ro}
\author{V.~Horny}
\affiliation{Extreme Light Infrastructure - Nuclear Physics, "Horia Hulubei" National Institute for R\&D in Physics and Nuclear Engineering, 30 Reactorului Street, 077125 Magurele, Romania}
\author{D.~Doria}
\affiliation{Extreme Light Infrastructure - Nuclear Physics, "Horia Hulubei" National Institute for R\&D in Physics and Nuclear Engineering, 30 Reactorului Street, 077125 Magurele, Romania}
\author{K.~Spohr}
\affiliation{Extreme Light Infrastructure - Nuclear Physics, "Horia Hulubei" National Institute for R\&D in Physics and Nuclear Engineering, 30 Reactorului Street, 077125 Magurele, Romania}
\affiliation{School of Computing, Engineering and Physical Sciences, University of the West of Scotland, High Street, PA1 2BE, Paisley, Scotland}


\begin{abstract}
Despite numerous achievements and recent progress, nuclear physics is often (wrongly) considered an old field of research nowadays. However, developments in theoretical frameworks and reliable experimental techniques have made the field mature enough to explore many new frontiers. In this regard, extending existing knowledge to an emerging field of physics---where particles interact with a relatively low-energy but high intensity field (intense enough so that multi-particle processes become comparable or more important than one-to-one processes)---can lead to exciting discoveries. Investigations can be realized under a highly time-compressed beam source (e.g., particle sources generated by laser-matter interaction using high-power laser systems). Here we focus on a new scheme, where high-power laser systems are exploited as a driver to generate energetic ($\gamma$-ray) photons. Together with additional low-energy photons provided by a second, less intense laser, a multi-photon absorption scheme enables a very attainable manipulation of nuclear transitions including isomer pumping and depletion. 
\end{abstract}

\maketitle

\section{Motivations}

As with many other physical phenomena, nuclear structure and dynamics involve quantum corrections to first-order processes that are often approached based on an expansion in terms of low energy or momentum ($\equiv M_{lo}$) over a high breakdown scale ($\equiv M_{hi}$). Since most quantum field theories are to be comprehended as effective field theories~\cite{Weinberg2021,de9}, higher-order corrections need to be generally suppressed by a positive power in $(M_{lo}/M_{hi})$ for a theory to have convergent results. 
Explicitly, up to few-body level, an observable evaluated up to order $n$ in general scales as~\cite{Griesshammer:2015osb,Griesshammer:2020fwr,yang_rev}: 
\begin{align}
\mathcal{O}_n(M_{lo};\Lambda;M_{hi})=\underbrace{\sum_i^n\left(\frac{M_{lo}}{M_{hi}}\right)^i\mathcal{F}_i(M_{lo};M_{hi})}_{\text{trustable part}} \\ \notag
+\underbrace{\operatorname{\mathscr C}_n(\Lambda;M_{lo},M_{hi})\left(\frac{M_{lo}}{M_{hi}}\right)^{n+1}}_{\text{uncertainty}},
\label{pc}
\end{align}
where $\Lambda$ is a cutoff where physics are truncated and renormalized, and the size of the explicit calculated function at order $n$, $\mathcal{F}_n$, must be in general comparable to the function $\operatorname{\mathscr C}_n$ so that the ``trustable part" can be improved order-by-order. 

This “$M_{lo}/M_{hi}$” hierarchy forbids rare events---which are higher-order effects---to be extracted clearly and studied straightforwardly from typical experimental data. Even probing with high-energy accelerators, where higher-order effects are enhanced (due to larger $M_{lo}$), as long as they are not dominating, underpinning new phenomena/theories involves complicated analysis and remains challenging. In short, there appears to be a dilemma toward probing new physics/phenomena with conventional accelerators.

However, certain higher-order corrections (for example, those involving multi-particle interactions) can become dominating due to a combinatorial enhancement~\cite{yang_rev,Yang:2021vxa}, especially under dense and/or intense conditions. 
The combinatorial enhancement occurs because:
\begin{itemize}
    \item There might exist intrinsic three- or higher-body forces.
    \item Under quantum theory, the uncertainty principle permits for more-to-one process even in the absence of higher-body forces.
\end{itemize}
Thus, instead of the one-to-one interaction, more-to-one processes are allowed. Moreover, in high-intensity environments, occurrence of these unconventional events grows combinatorially, dominates the observables, and enables straightforward studies.  

\begin{figure}[h]
  \includegraphics[width=0.9\linewidth]{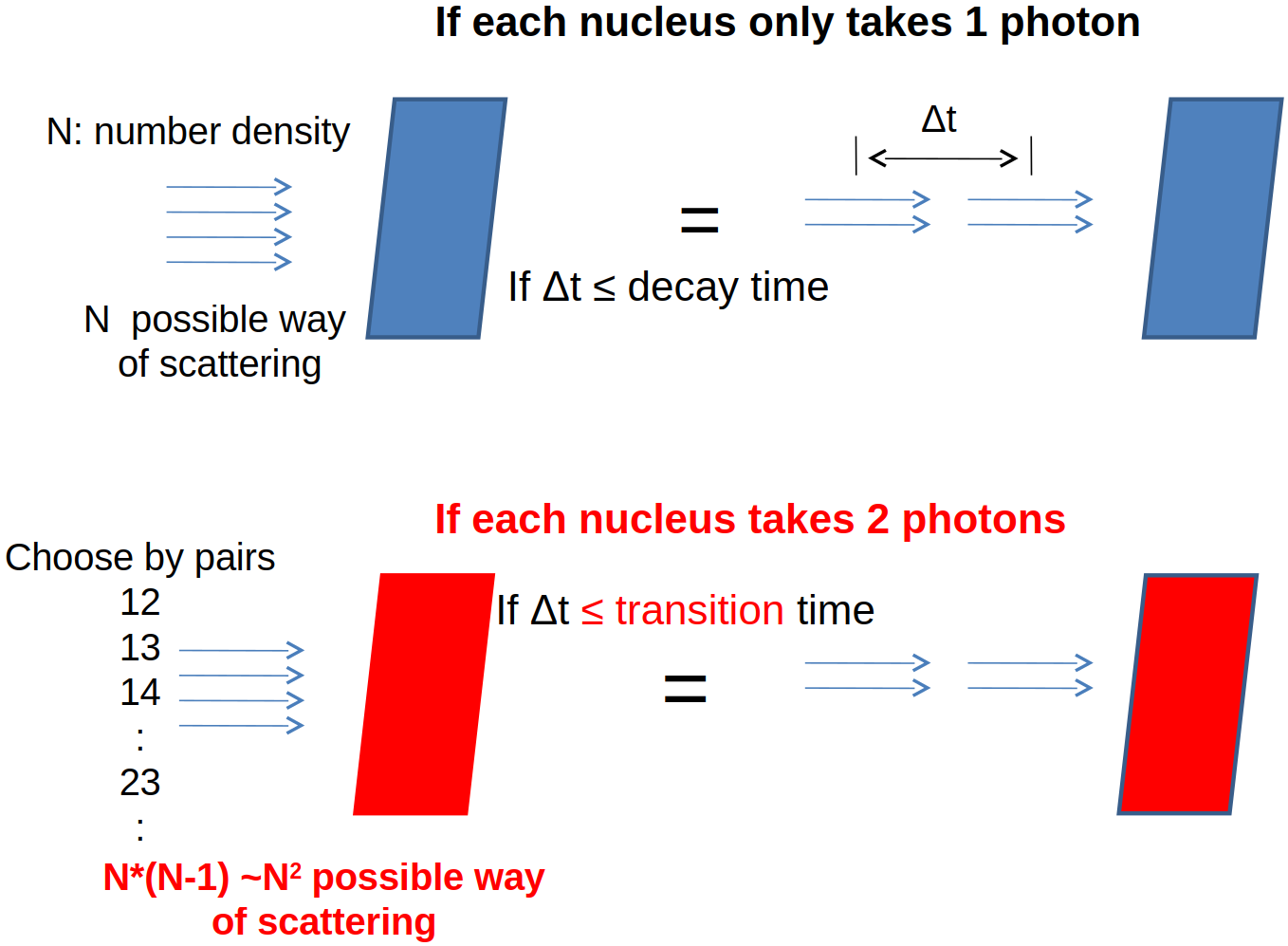}
  \caption{Illustration of the origin of non-linearity in two-photon absorption process. The blue arrows denote the number density of incoming photons, which interact with atomic/nucleus target.}
  \label{fig1}
\end{figure}

Multi-photon absorption process is a premier example of combinatorial enhancement, which is illustrated in Fig.~\ref{fig1}. The upper panel (illustrated with blue targets) describes conventional stepwise photon absorption, where each nucleus in the target interacts with one photon per time. For a photo flux corresponding to $N$ incoming photons per unit area per unit time, there are ``$N$ possible ways of scattering" to excite each nucleus. The yield is straightforwardly proportional to the intensity of the photon beam. Meanwhile, the lower panel (illustrated with red targets) describes the two-photon-absorption (2PA) process~\cite{GM}. For 2PA, each excitation involves two photons. This process is a second-order effect on the transition operator and generally comes with a smaller cross section. However, in contrast to stepwise one-photon absorption, there are $N(N-1)\sim N^2$ combinations of two-photon pairs. This means that the combinatorial factor will overcome the small cross section for $N\gg1$, leading to a yield that exceeds 1PA. This is the very origin of the non-linearity of two- or multi-photon absorption (nPA).

The above combinatorial enhancement has been demonstrated experimentally in non-linear optics~\cite{PhysRevLett.7.229,bloembergen1992nonlinear}. Under intense optical-photon flux provided by lasers, multi-photon absorption---which belongs to higher-order effects on the interaction Hamiltonian---has been shown to dominate over conventional single-photon processes in the atomic/molecule transitions~\cite{Protopapas1997}. Interestingly, the same is yet to be observed in the nuclear case due to the lack of intensive $\gamma$-ray sources. Nevertheless, its realization could have invaluable applications/impacts in nuclear physics and astrophysics. 
In the following, we argue that recent breakthroughs in high-power laser systems (HPLS) provide a unique opportunity that could lead nuclear/particle physics to a new forefront of scientific exploration.

\begin{figure}[h]
  \includegraphics[width=0.9\linewidth]{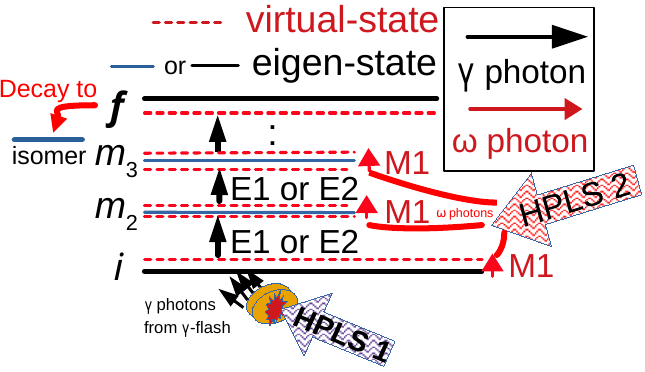}
  \caption{Illustration of the scheme, where the pumping is achieved through multi-photon absorption with transitions across each virtual state completed by a pair of $\gamma +\omega$ photon. The $\gamma$-photons carry out E1 or E2 transitions and are generated via laser-matter interaction. The main enhancement is achieved through the $\omega$-photons, which conduct M1 transitions and are directly supplied by a second HPLS with a duration covering the entire period of the $\gamma$-flash. }
  \label{fig}
\end{figure}


\section{Main Concepts}

It has been demonstrated that shooting a high-power laser pulse onto an over-dense target could generate dense $\gamma$-photons that present a continuous energy spectrum~\cite{PhysRevLett.108.165006,Ji2014}. 
However, the intensity of $\gamma$-photons achievable today has not yet reached the level for 2PA to be carried through solely from $\gamma$-photons~\cite{our_iso}.  
Fortunately, the photons that participate in 2PA need not to be of the same energy.
The theoretical derivation of 2PA was first introduced by G\"{o}ppert-Mayer~\cite{GM} with the following key formula:
\begin{equation}
\frac{d\mathcal{Y}_{(2)}}{dz}=\mathcal{N}\cdot(I_1I_2\sigma_{2pa}),\label{eq1}
\end{equation}
where $\mathcal{Y}_{(2)}$ denotes the yields of the final state per unit time given by 2PA, $I_i$ is the number intensity of photons in the $i^{th}$-category, 
$\mathcal{N}$ is the number of nuclei photons encountered per distance $dz$ in the target, and 
$\sigma_{2pa}$ is the generalized cross-section for 2PA. In contrast to conventional 1PA, where 
\begin{equation}
\frac{d\mathcal{Y}_{(1)}}{dz}=\mathcal{N}\cdot(I_1\sigma_{1pa}),\label{eq2}
\end{equation}
the yield from Eq.~(\ref{eq1}) depends on both $I_1$ and $I_2$. Since both $\sigma_{1pa}$ and $\sigma_{2pa}$ have a fixed value for each selected transition, as $I_2$ grows, the contribution of $I_1I_2\sigma_{2pa}$ will eventually \textit{exceed} $I_1\sigma_{1pa}$, leading to a controllable enhancement depending on $I_2$. In other words, one can consider the scenario that $I_1$ represents the intensity of $\gamma$-photons, while $I_2$ is supplied by optical photons ($\equiv \omega$-photons) which come directly from a laser.   
\begin{figure}[h]
\includegraphics[width=0.9\linewidth]{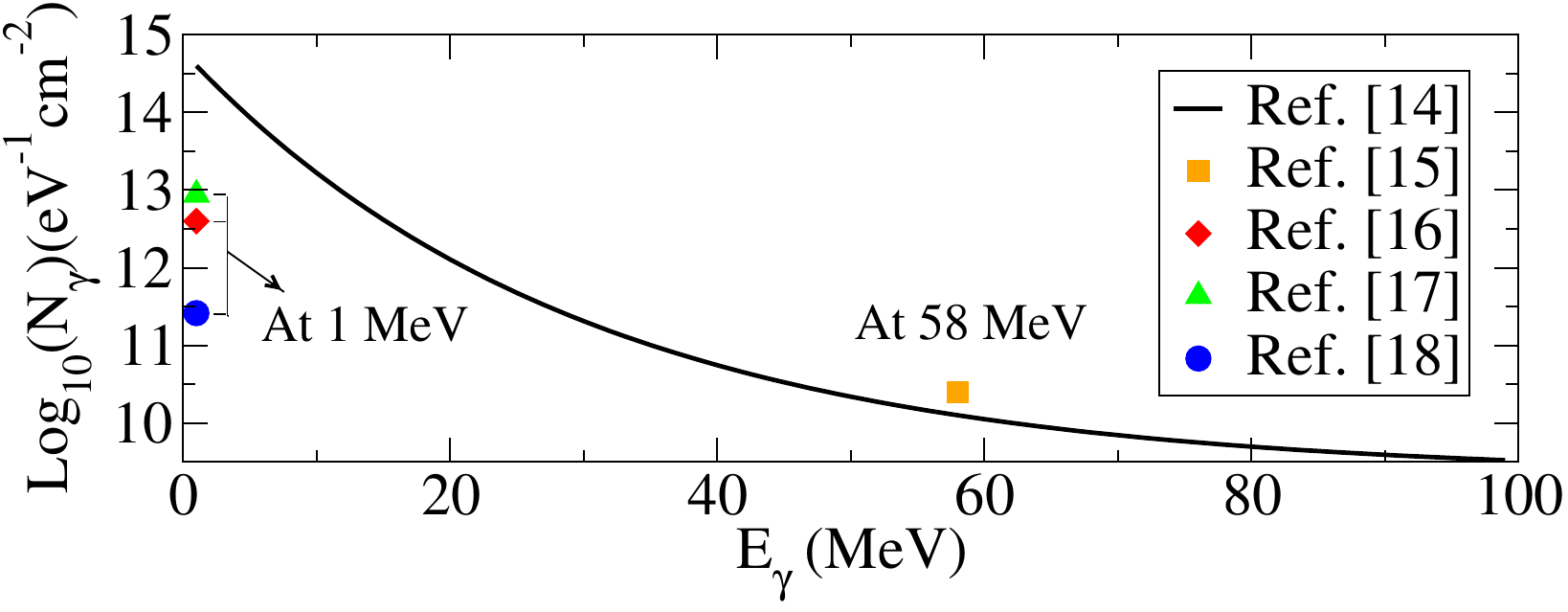}
\caption{Number of $\gamma$-photons N$_{\gamma}$ per eV cm$^2$ as a function of energy generated via laser-plasma interaction in a single shot converted from Refs.~\cite{PhysRevApplied.13.054024,Gu2018,Xue2020,PhysRevE.104.045206,Heppe2022}. Note that the actual $\gamma$-flux occupies an area of only $\approx 100$ $\mu m^2$. The accumulated time is $t\approx 15-50$ fs. The figure is adopted from Ref.~\cite{our_iso} with conversion details discussed therein.}
\label{fig2}
\end{figure}

Refs.~\cite{PhysRevLett.42.1397,PhysRevC.20.1942} have evaluated $\sigma^{2pa}_{eff}\equiv I_2\sigma_{2pa}$, i.e., the effective cross section seen by photons in the $I_1$-category and found that the yield from 2PA dominates over 1PA when photons in the $I_2$-category are supplied by a $10^{10}$ W/cm$^{2}$ laser. Modern HPLS can generate $\omega$-photons that were far beyond reach, therefore allowing one to go beyond 2PA and achieve efficient pumping/depopulation of nuclear isomers through the more general n-photon process (nPP)---where one considers any combination of absorption and emission of n photons at once~\cite{Delone2000}.
A possible scheme is illustrated in Fig.~\ref{fig}, where HPLS are exploited as a driver to generate energetic ($\gamma$-ray) photons. Together with additional low-energy photons provided by a second, less intense laser, the nPP scheme is evaluated in terms of achievable cross sections and corresponding isomer population numbers per laser shot~\cite{our_iso}, resulting in the prediction of a drastic increase by non-linear yield enhancement in multi-photon absorption processes. With the typical intensity of the $\gamma$ flash given by particle-in-cell (PIC) simulations as listed in Fig.~\ref{fig2} (via laser-matter interaction driven by PW-class HPLS1) and $\approx 10^{18}$ W/cm$^{2}$ supplied photons ($\omega$-photons given by HPLS2), a 10$^{10}$-fold improvement is expected against all previous proposals in isomer pumping/depletion\footnote{For isomers separated from their ground state within E4 transitions.}.

\section{Experimental Design/Setup}

The experimental setup of the above isomer pumping/depletion scheme (as illustrated in Fig.~\ref{fig}) turns out to be surprisingly simple. Since nPP will be maximized whenever the virtual states approach the physical intermediate states, $\omega$-photons (supplied by HPLS2) will automatically select and combine with $\gamma$-photons with suitable energies. The continuous energy spectrum generated by the laser-matter interaction guarantees that intense $\gamma$-rays with 10 keV $\lesssim E_{\gamma}\lesssim 10$ MeV are available to accommodate various transition gaps through virtual states toward the desired state in most nuclei. 

Practically, only two photon sources are required: an ultra-short PW-class laser pulse (HPLS1) to generate $\gamma$-photons via laser-matter interactions, and another (HPLS2) to supply $\omega$-photons.

\section{Calculations based on a real nuclear system}
As explained above, due to the continuous spectrum of HPLS-generated $\gamma$-rays, the aforementioned scheme can be applied to a very wide range of nuclei. To get a general idea of the magnitude of nPP, Refs.~\cite{PhysRevLett.42.1397,PhysRevC.20.1942,our_iso,Yang:2024chv} adopted the Weisskopf formula to calculate nuclear transitions. The inaccuracy/error due to the single-particle approximation can be up to $\times/\div100$ times compared to the experiments, large-scale shell-model based calculations, or any alternatives, but still within a tolerance compensable by supplying more intense $\omega$-photons.

\begin{figure}[h]
\includegraphics[width=0.9\linewidth]{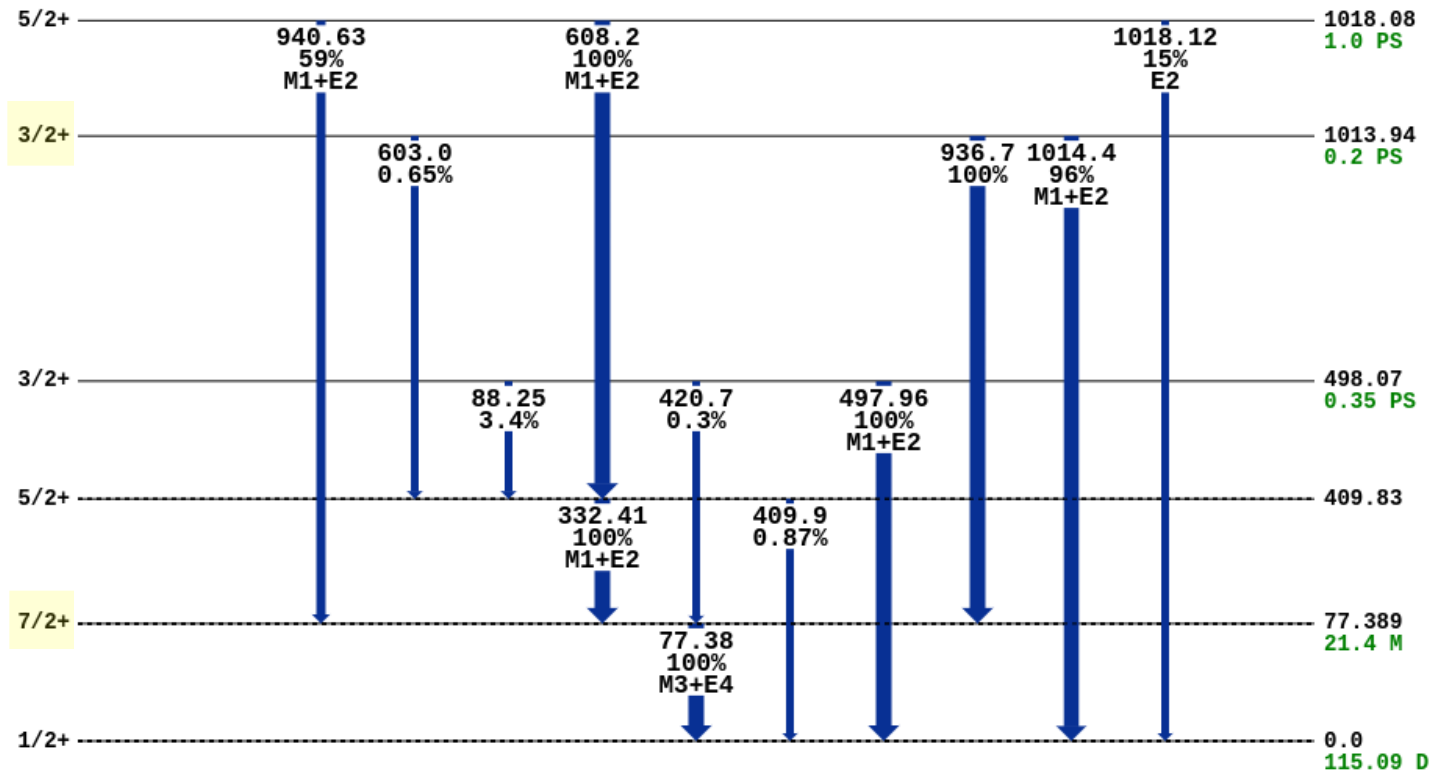}
\caption{Level scheme of $^{113}$Sn adopted from https://www.nndc.bnl.gov.}
\label{113sn}
\end{figure}

It is crucial to probe a few special cases where the multi-photon transitions are massively enhanced. However, this requires sophisticated nuclear structure calculations because most transitions between intermediate states with half-life $t_{1/2}\ll 1$ ps are not well determined experimentally. 
Here, an example of an enhanced transition from the $J^{\pi}=\frac{7}{2}^+$ (77.389 keV) isomer state of the $^{113}$Sn nucleus through 2PP is provided to demonstrate the principle. The $J^{\pi}=\frac{7}{2}^+$ (77.389 keV) state has $t_{1/2}=21.4$ minutes, which can absorb a pair of $\gamma+\omega$ photons to reach the $J^{\pi}=\frac{3}{2}^+$ (1013.94 keV) state ($t_{1/2}=0.2$ ps) and rapidly decays to the ground state. The relevant level scheme is illustrated in Fig.~\ref{113sn}.

\begin{figure}[h]
\includegraphics[width=0.9\linewidth]{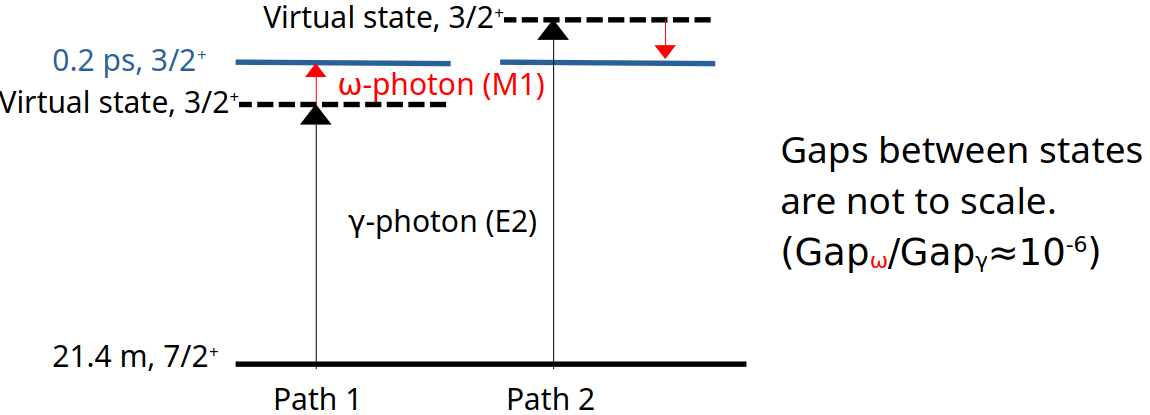}
\caption{Illustration of a transition enhanced by 2PP from the two highlighted states ($J^{\pi}=\frac{7}{2}^+$ to $J^{\pi}=\frac{3}{2}^+$) in Fig.~\ref{113sn}. Note that the energy of the $\omega$-photon is $\approx 10^{-6}$ of the energy of the $\gamma$-photon.}
  \label{fig_dia_2pp}
\end{figure}

Since $t_{1/2}$ and the decay branching ratios of the $\frac{3}{2}^+$ state are experimentally available, the transition amplitudes can be obtained directly. I.e., the transition rate from a state $a$ to $b$ is
\begin{align}
\lambda_{a\rightarrow b}=1/\tau_{a\rightarrow b},
 \label{eqma}
\end{align}
where the mean life $\tau$ is related to half-life by $\tau=t_{1/2}/\ln(2)$.
For the highlighted states, the partial mean life $\tau_{\frac{3}{2}^+\rightarrow\frac{7}{2}^+}$ is
\begin{align}
    \frac{1}{\tau_{\frac{3}{2}^+\rightarrow\frac{7}{2}^+}}=\frac{100}{196.65}\frac{1}{\tau_{\frac{3}{2}^+\rightarrow all}}.
\end{align}
The 2PA transition between the highlighted states assisted by the $\omega$-photon can then be achieved by the process described in Fig.~\ref{fig_dia_2pp}.

Here, the virtual state 2PP went through has the same structure (i.e., same $J^{\pi}$ and almost identical wavefunction) as the 0.2 ps, $\frac{3}{2}^+$ state, with the only difference being that it is slightly separated from the $\frac{3}{2}^+$ state by $E_{\omega}$.
Assuming that the supplied $\omega$-photons have $E_{\omega}\approx 1$ eV and a bandwidth$\approx 1$ eV, and are overlapped with the HPLS-generated $\gamma$-photons with $E_{\gamma}=936.7\text{ keV}-E_{\omega}\pm (\text{bandwidth})/2$, 2PA is then achieved by Path 1 (Path 2 is achieved by replacing $-E_{\omega}$ with $+E_{\omega}$). The transition matrix element carried out by the $\gamma$-photon can be obtained directly from data through Eq.~(\ref{eqma}), and the transition matrix element of the $\omega$-photon from the virtual state to the $\frac{3}{2}^+$ state can be described nearly exactly by the M1 transition under the Weisskopf estimate. 
Explicit calculations show that the effective cross section of 2PA, i.e., $\sigma^{\mathrm{2PA}}_\mathrm{eff}\approx 5.2\times 10^{-48}\mathcal{P}_{\omega}$ cm$^2$\footnote{The total 2PP cross section by adding up path 1 and 2 is then approximately twice of $\sigma^{\mathrm{2PA}}_\mathrm{eff}$.}. In other words, the $\gamma$-photons that participate in 2PP will be absorbed with a cross section $\sigma^{\mathrm{2PP}}_\mathrm{eff}\approx 10^{-25}$ cm$^2$ if one overlaps them with $\omega$-photons supplied by a PW-class ($\mathcal{P}_{\omega}\approx 10^{22}$ W/cm$^2$) laser. This extraordinary cross section (0.1 barn) indicates that an originally rare transition via 1-photon absorption can be made possible and even very large by the non-linear mechanism of 2PP.  
Moreover, transitions that go through $t_{1/2}\approx1-10$ fs states---though not well measured experimentally---are of great abundance and generally correspond to much larger transition amplitudes (as $\lambda=ln(2)/t_{1/2}$). The above 2PP transition (utilizing the $t_{1/2}=0.2$ ps state) will be further enhanced if a $t_{1/2}\approx1$ fs intermediate state is adopted (cf. Fig.~1 of Ref.~\cite{PhysRevC.20.1942}), which then makes the nPP scheme (with n$>2$, as illustrated in Fig.~\ref{fig}) feasible and therefore enables efficient pumping and depletion of isomers~\cite{our_iso}. 

\section{Future Prospects}
The above scheme can be applied to pump and deplete a very wide class of nuclear isomers. Therefore, it represents a crucial step toward the realization of exciting new concepts in nuclear photonics. Furthermore, the technique of enhancing effective cross section by overlapping the $\gamma$- and $\omega$-photons given by HPLS can potentially be utilized to uncover hidden states and rare transitions in the nuclei.
Moreover, it provides a novel way to achieve the stimulated amplification of $\gamma $-rays (graser). In Ref.~\cite{Yang:2024chv}, it is shown that the so-called graser dilemma, which originated from Einstein’s theory utilizing detailed balancing~\cite{1916DPhyG..18..318E}, can be circumvented through a non-linear multi-photon mechanism. The supplied $\omega$-photons also provide venues to overcome the problems with respect to recoil and broadening of the narrow absorption line breadth. As a result, calculations suggest that a graser based on selected nuclear isomers could be realized already on those sites equipped with PW-class high-power laser systems. 

Further theoretical and experimental studies of multi-photon mechanism utilizing HPLS are in progress. To accelerate progress in this transformative paradigm, innovation and multi-discipline collaboration between researchers in nuclear and laser-plasma communities are required to unlock the immense possibilities that lie ahead.

\acknowledgments 
 
We thank C. Bertulani and P. Tomassini for
useful discussions and suggestions. This work was supported by the
the Extreme Light Infrastructure Nuclear Physics (ELI-NP) Phase II, a project co-financed by the Romanian Government and the European Union through the European Regional Development Fund - the Competitiveness Operational Programme (1/07.07.2016, COP, ID 1334);  the Romanian Ministry of Research and Innovation: PN23210105 (Phase 2, the Program Nucleu); and ELI-RO\_RDI\_2024\_AMAP, ELI-RO\_RDI\_2024\_LaLuThe, ELI-RO\_RDI\_2024\_SPARC of the Romanian Government.
We acknowledge PRACE for awarding us access to Karolina at IT4Innovations, Czechia under project number EHPC-BEN-2023B05-023 (DD-23-83); IT4Innovations at Czech National Supercomputing Center under project number OPEN24-21 1892; Ministry of Education, Youth and Sports of the Czech Republic through the e-INFRA CZ (ID:90140) and CINECA under PRACE EHPC-BEN-2023B05-023.

\bibliography{main} 
\end{document}